\documentclass[aps,prb,twocolumn,superscriptaddress,showpacs]{revtex4}
\pdfoutput=1
\usepackage{array}
\usepackage{graphicx}
\usepackage{amssymb}

\begin{document}

\title{Neutron powder diffraction study on the non-superconducting phases of ThFeAsN$_{1-x}$O$_x$ ($x=0.15, 0.6$) iron pnictide}

\author{Huican Mao}
\affiliation{Department of Physics and Center for Advanced Quantum Studies, Beijing Normal University, Beijing 100875, China}
\affiliation{Beijing National Laboratory for Condensed Matter
Physics, Institute of Physics, Chinese Academy of Sciences, Beijing
100190, China}
\author{Bingfeng Hu}
\email{hbf@caep.cn}
\affiliation{Key Laboratory of Neutron Physics, Institute of Nuclear Physics and Chemistry, China Academy of Engineering Physics, Mianyang 621999, China}
\author{Yuanhua Xia}
\affiliation{Key Laboratory of Neutron Physics, Institute of Nuclear Physics and Chemistry, China Academy of Engineering Physics, Mianyang 621999, China}
\author{Xiping Chen}
\affiliation{Key Laboratory of Neutron Physics, Institute of Nuclear Physics and Chemistry, China Academy of Engineering Physics, Mianyang 621999, China}
\author{Cao Wang}
\affiliation{Department of Physics, Shandong University of Technology, Zibo 255049, China}
\author{Zhicheng Wang}
\affiliation{Department of Physics and State Key Lab of Silicon Materials, Zhejiang University,
Hangzhou 310027, China}
\author{Guanghan Cao}
\affiliation{Department of Physics and State Key Lab of Silicon Materials, Zhejiang University,
Hangzhou 310027, China}
\affiliation{Collaborative Innovation Centre of Advanced Microstructures, Nanjing 210093, China}
\author{Shiliang Li}
\affiliation{Beijing National Laboratory for Condensed Matter
Physics, Institute of Physics, Chinese Academy of Sciences, Beijing
100190, China}
\affiliation{Songshan Lake Materials Laboratory, Dongguan, Guangdong 523808, China }
\affiliation{School of Physical Sciences, University of Chinese Academy of Sciences, Beijing 100190, China}
\author{Huiqian Luo}
\email{hqluo@iphy.ac.cn}
\affiliation{Beijing National Laboratory for Condensed Matter
Physics, Institute of Physics, Chinese Academy of Sciences, Beijing
100190, China}
\affiliation{Songshan Lake Materials Laboratory, Dongguan, Guangdong 523808, China }

\begin{abstract}
We use neutron powder diffraction to study on the non-superconducting phases of ThFeAsN$_{1-x}$O$_x$ with $x=0.15, 0.6$. In our previous results on the superconducting phase ThFeAsN with $T_c=$ 30 K, no magnetic transition is observed by cooling down to 6 K, and possible oxygen occupancy at the nitrogen site is shown in the refinement(H. C. Mao \emph{et al.}, EPL, 117, 57005 (2017)). Here, in the oxygen doped system ThFeAsN$_{1-x}$O$_x$, two superconducting region ($0\leqslant x \leqslant 0.1$ and $0.25\leqslant x \leqslant 0.55$) have been identified by transport experiments (B. Z. Li \emph{et al.}, J. Phys.: Condens. Matter 30, 255602 (2018)). However, within the resolution of our neutron powder diffraction experiment, neither the intermediate doping $x=0.15$ nor the heavily overdoped compound $x= 0.6$ shows any magnetic order from 300 K to 4 K. Therefore, while it shares the common phenomenon of two superconducting domes as most of 1111-type iron-based superconductors, the magnetically ordered parent compound may not exist in this nitride family.
\end{abstract}


\maketitle
Understanding the magnetic ground state is one of the major task in the mechanism research of iron-based superconductors, where the magnetic interactions are generally believed to be involved in the superconducting pair process \cite{pdai2015,si2016,inosov2016}. In most families of iron pnictide or chalcogenide superconductors, a long-range antiferromagnetic order always emerges in the undoped parent compounds \cite{stewart2011}, such as the collinear order in LaFeAsO (1111 family) \cite{cruz2008}, BaFe$_2$As$_2$(122 family) \cite{qhuang2008}, Na$_{1-\delta}$FeAs(111 family) \cite{slli2009a}, the bi-collinear order in Fe$_{1+x}$Te(11 family) \cite{slli2009b}, and the $\sqrt{5}\times\sqrt{5}$ block order in K$_{2}$Fe$_{4}$Se$_5$ \cite{wbao2011,fye2011,mwang2011}. Even in the stoichiometrically hole-type superconducting system CaKFe$_4$As$_4$ (1144 family), compensating electron doping by Ni or Co can induce a spin-vortex phase (hedgehog order) with C$_4$ rotation symmetry \cite{wmeier2018,akreyssig2018}. Another C$_4$ magnetic order can be found in the hole-doped Ba$_{1-x}$(K, Na)$_x$Fe$_2$As$_2$ systems near the optimal doping \cite{aebohmer2015,savci2014}, too. Specifically for the 1111 family (e.g. LaFeAsO) \cite{hosono2015,kamihara2008}, superconductivity can be induced by doping fluorine \cite{kamihara2008,hluetkens2009,jyang2015,jyang2018,yxiao2008}, hydrogen \cite{smatsuishi2012,siimura2012,nfujiwara2013,hhosono2013} and phosphorus \cite{cwang2009,skitagawa2014,smiyasaka2013,ktlai2014,cshen2016,smiyasaka2017} into the parent compound, where all of them result in a two superconducting dome structure with slightly different optimal $T_c$. Antiferromagnetic parent compounds for both side are discovered by neutron diffraction, muon spin relaxation ($\mu$SR) and nuclear magnetic resonance (NMR) experiments \cite{nfujiwara2013,hhosono2013,hmukuda2014,mhiraishi2014}.

Recently, a new nitride iron pnictide superconductor ThFeAsN has been discovered with intrinsic $T_c=$ 30 K \cite{cwang2016}. The layered tetragonal ZrCuSiAs-type structure consisting of [Th$_2$N$_2$] and [Fe$_2$As$_2$] blocks is classified as a 1111-type iron-based superconducting family (Insert of Fig. 1). Although the first-principles calculations of ThFeAsN indicate that the lowest-energy magnetic ground state is the collinear stripe-type antiferromagnetic state \cite{djsingh2016,gwang2016}, the normal-state resistivity / magnetization do not show any magnetic anomaly above $T_c$ \cite{cwang2016}, further $^{57}$Fe M\"{o}ssbauer spectroscopy, neutron powder diffraction, $\mu$SR and NMR experiments have not found any magnetic order down to 2 K \cite{maalbedah2017,hcmao2017,tshiroka2017,dadroja2017}, either. Possible extra electrons from nitrogen deficiency or oxygen occupancy at the nitrogen site make the compound approximating to the optimal doping level \cite{hcmao2017}. Indeed, by doping oxygen into the system, the superconductivity of ThFeAsN$_{1-x}$O$_x$ is quickly suppressed until $x=0.1$ \cite{bzli2018}. Surprisingly, a second superconducting dome emerges from $x=0.25$ to $x=0.55$ with maximum $T_c$ about 15 K (Fig. 1) \cite{bzli2018}, which closely resembles the two superconducting domes in the phase diagram of LaFeAsO$_{1-x}$H$_x$, LaFeAsO$_{1-x}$F$_x$ and LaFeAs$_{1-x}$P$_x$O systems \cite{jyang2015,smatsuishi2012,siimura2012,nfujiwara2013,hhosono2013,cwang2009,skitagawa2014,smiyasaka2013,ktlai2014,smiyasaka2017}. It should be noticed that the hydrogen doped 1111 systems always have double parent compounds with different magnetic structure and ordered moment adjacent to each superconducting dome \cite{nfujiwara2013,hhosono2013,mhiraishi2014}, and the intermediate non-superconducting phase of LaFeAs$_{1-x}$P$_x$O is magnetically ordered \cite{cshen2016,hmukuda2014,smiyasaka2017}. Thus it is essential to examine the possible magnetic order in ThFeAsN$_{1-x}$O$_x$ system, especially for the non-superconducting dopings.

\begin{figure}[h]
\includegraphics[width=0.4\textwidth]{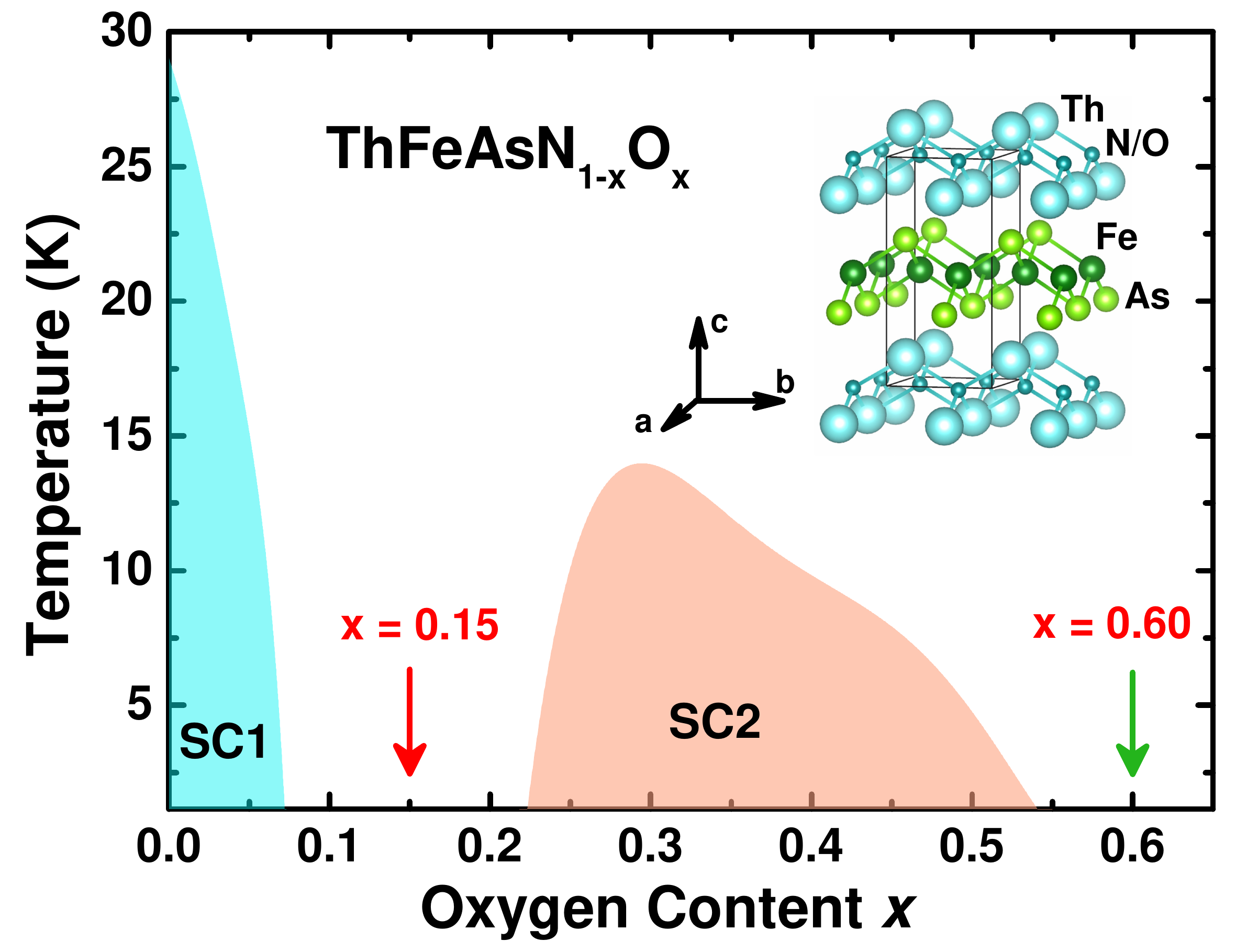}
\caption{Superconducting phase diagram and crystal structure of ThFeAsN$_{1-x}$O$_x$. The arrows mark two dopings in this study.
 }
\end{figure}

In this letter, we report a neutron powder diffraction study on the non-superconducting phases of ThFeAsN$_{1-x}$O$_x$ with $x=0.15, 0.6$. By cooling down from 300 K to 4 K, no magnetic order is found within the instrument resolution. The refinement shows a slightly compression of the unit cell volume and a lift of the Th position upon oxygen doping. Together with our previous results on the undoped compound ThFeAsN, we conclude that no magnetic parent compound exist in this nitride family. The two separated superconducting region probably is a combination effect from electron doping and uniaxial chemical pressure from oxygen substitutions.

\begin{center}
\begin{figure}[t]
\includegraphics[width=0.4\textwidth]{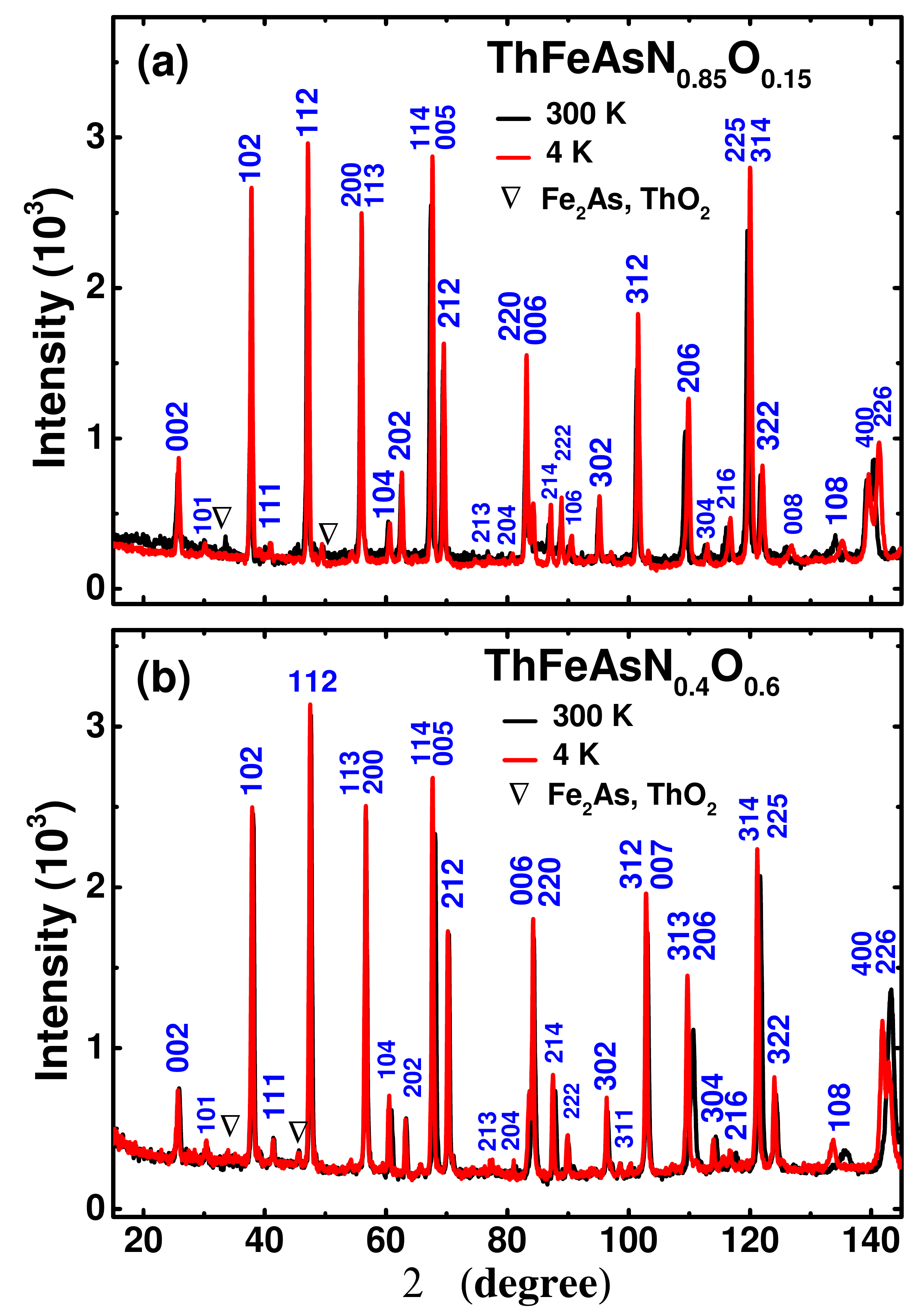}
\caption{Raw data of neutron powder diffraction for (a) ThFeAsN$_{0.85}$O$_{0.15}$ and (b) ThFeAsN$_{0.4}$O$_{0.6}$. All peaks are indexed by the tetragonal ZrCuSiAs-type structure.
 }
\end{figure}
\end{center}

Polycrystalline samples of ThFeAsN$_{1-x}$O$_x$ ($x=0.15, 0.6$) were synthesized by the solid-state reaction method as described elsewhere \cite{cwang2016,bzli2018}. About 5 grams high pure powders were prepared for each doping, and sealed in a vanadium can. Neutron powder diffraction experiments were carried out on HRND (High Resolution Neutron Diffractometer) at the Key Laboratory of Neutron Physics, Institute of Nuclear Physics and Chemistry, China Academy of Engineering Physics. The wavelength of neutron was selected to be $\lambda$ = 1.8846 {\AA} for both samples. The scattering data was collected at 4 K and 300 K by covering the scattering angle $2\theta$ range 10 - 145 degrees. All these diffraction patterns were refined with Rietveld method within the program FullProf \cite{jrc1993}, and the structural parameters were obtained by assuming the occupancy same as chemical composition \cite{hcmao2017}.

\begin{figure*}[t]
\includegraphics[width=0.75\textwidth]{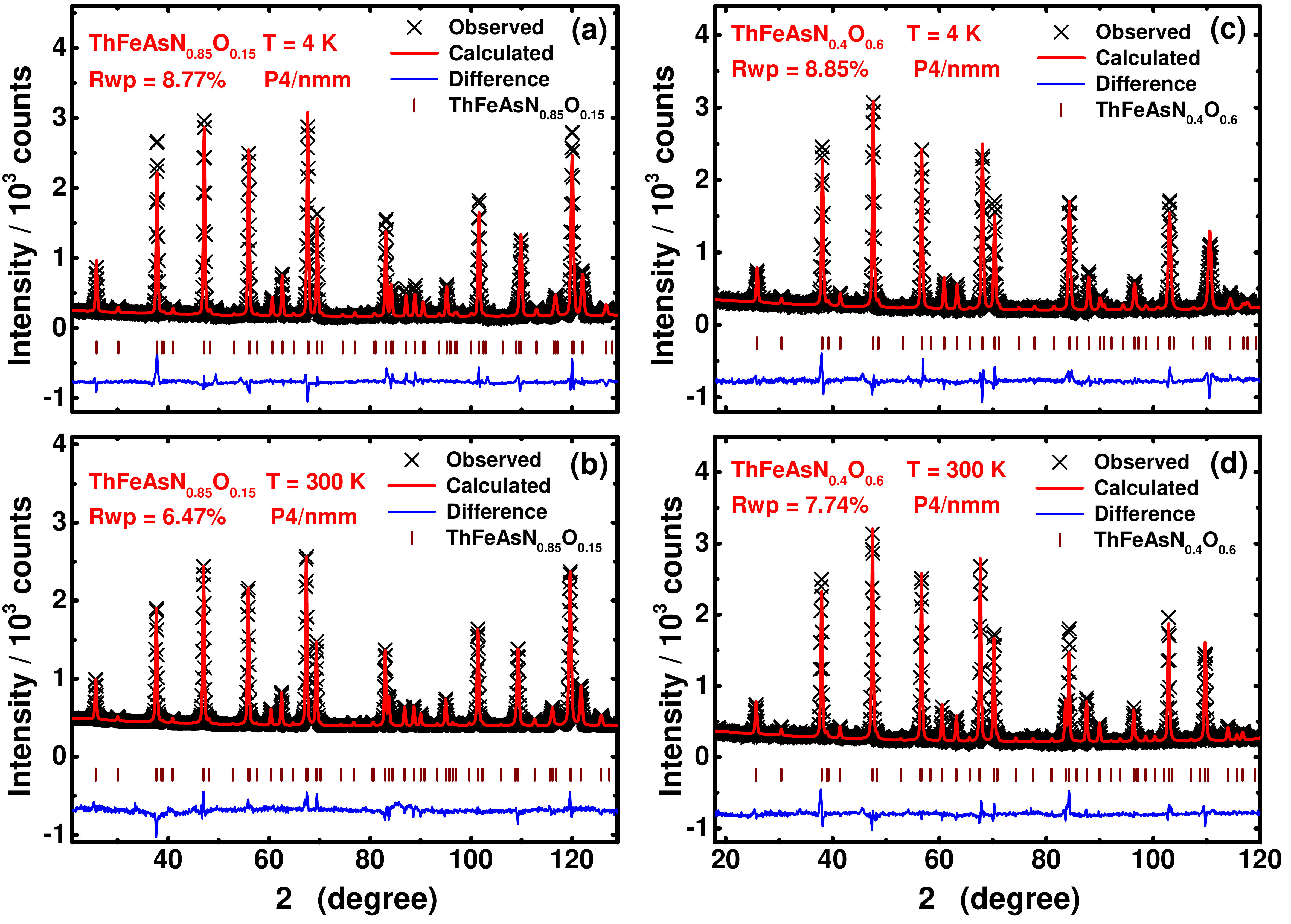}
\caption{(a) and (b): Refinement results of neutron powder diffraction patterns for ThFeAsN$_{0.85}$O$_{0.15}$ at 4 K and 300 K. (c) and (d): Identical refinement results for ThFeAsN$_{0.4}$O$_{0.6}$ at 4 K and 300 K.
 }\label{fig3}
\end{figure*}

The raw data of neutron diffraction patterns are presented in Fig. 2. For the $x=0.15$ sample, the diffraction patterns are almost overlap between $T=4 $ K and $T=300 $ K. All reflections can be indexed by a tetragonal phase in ZrCuSiAs-type structure \cite{smuir2012} with the space group $P4/nmm$, except for two tiny peaks (marked by inverted triangles) from Fe$_2$As or ThO$_{2}$ impurity phases (Fig. 2(a)). Similar results are obtained on the $x=0.6$ sample, as shown in Fig. 2(b). As both data sets at base temperature and room temperature can be fully identified as nuclear peaks, and no additional peaks emerge at low temperature, we tend to consider no magnetic orders in both compounds within the measurement resolution of HRND.  Here the slight difference between $T=4 $ K and $T=300 $ K patterns can be attributed to the background change, and the peak shift at high angles are due to thermal expansion of the lattice.

To quantitatively compare the neutron diffraction results, we have performed the Rietveld refinement for four data sets by assuming the occupancy same as chemical composition, as shown in Fig. 3. The weighted profile factor Rwp without the background is slightly higher than previous results on ThFeAsN \cite{hcmao2017}, which may be attributed to the much lower neutron flux of this high resolution diffractometer in comparison with WOMBAT high-intensity diffractometer at the Australian Centre for Neutron Scattering (ACNS). Again, four sets of the diffraction patterns can be refined quite well by the space group $P4/nmm$ without any magnetic scattering, suggesting no static magnetic orders in both samples. This is consistent with the magnetic susceptibility measurements, where no anomaly related to magnetic transition can be observed in $\chi(T)$, and only Pauli paramagnetic behavior shows up instead of the Curie-Weiss behavior. All crystallographic parameters obtained from the refinements are listed in Table.1 and Table.2.

\begin{table}
\caption{Crystallographic data of ThFeAsN$_{0.85}$O$_{0.15}$ at 4 K.}
\label{Tab.1}
\begin{centering}
{
\begin{tabular}{lccr}
\hline
space group & P4/$nmm$     & R$wp$(\%)    & 8.77(3) \\
$a$(\AA)       & 4.0156(3)      &$h_{Th-As}$(\AA)  & 1.7037(8)\\
$c$(\AA)       & 8.4305(1)    & $h_{Fe-As}$(\AA)   &1.3185(1)\\
$\alpha$$_{As-Fe-As}$ &$113.4^{\circ}$  &$d_{Th-As}$(\AA) &3.311(4)\\
$\beta$$_{As-Fe-As}$ &$107.55^{\circ}$  &$d_{Fe-As}$(\AA) &2.403(3)\\
\end{tabular}
}
{
\begin{tabular}{lccccr}
\hline
atom     &Wyckoff    &x   &y   &z   &U$_{iso}$\\
Th       &2c &0.25 &0.25 &0.1415(1) &0.1232\\
Fe       &2b &0.75 &0.25 &0.5 &0.2542\\
As       &2c &0.25 &0.25 &0.6564(8) &0.1495\\
N        &2a &0.75 &0.25 &0 &0.3374\\
O        &2a &0.75 &0.25 &0 &0.3374\\
\hline
\end{tabular}
}

\end{centering}
\end{table}

\begin{table}
\caption{Crystallographic data of ThFeAsN$_{0.4}$O$_{0.6}$ at 4 K.}
\label{Tab.2}
\begin{centering}
{
\begin{tabular}{lccr}
\hline
space group & P4/$nmm$     & R$wp$(\%)    & 8.85(1) \\
$a$(\AA)       & 3.9727(9)      &$h_{Th-As}$(\AA)  & 1.5671(1)\\
$c$(\AA)       & 8.4253(1)    & $h_{Fe-As}$(\AA)   &1.4061(8)\\
$\alpha$$_{As-Fe-As}$ &$109.4^{\circ}$  &$d_{Th-As}$(\AA) &3.216(3)\\
$\beta$$_{As-Fe-As}$ &$109.5^{\circ}$  &$d_{Fe-As}$(\AA) &2.434(3)\\
\end{tabular}
}
{
\begin{tabular}{lccccr}
\hline
atom     &Wyckoff    &x   &y   &z   &U$_{iso}$\\
Th       &2c &0.25 &0.25 &0.1471(8) &1.0259\\
Fe       &2b &0.75 &0.25 &0.5 &0.8284\\
As       &2c &0.25 &0.25 &0.6669(3) &0.0694\\
N        &2a &0.75 &0.25 &0 &0.8296\\
O        &2a &0.75 &0.25 &0 &0.8296\\
\hline
\end{tabular}
}

\end{centering}
\end{table}

\begin{figure}[t]
\includegraphics[width=0.47\textwidth]{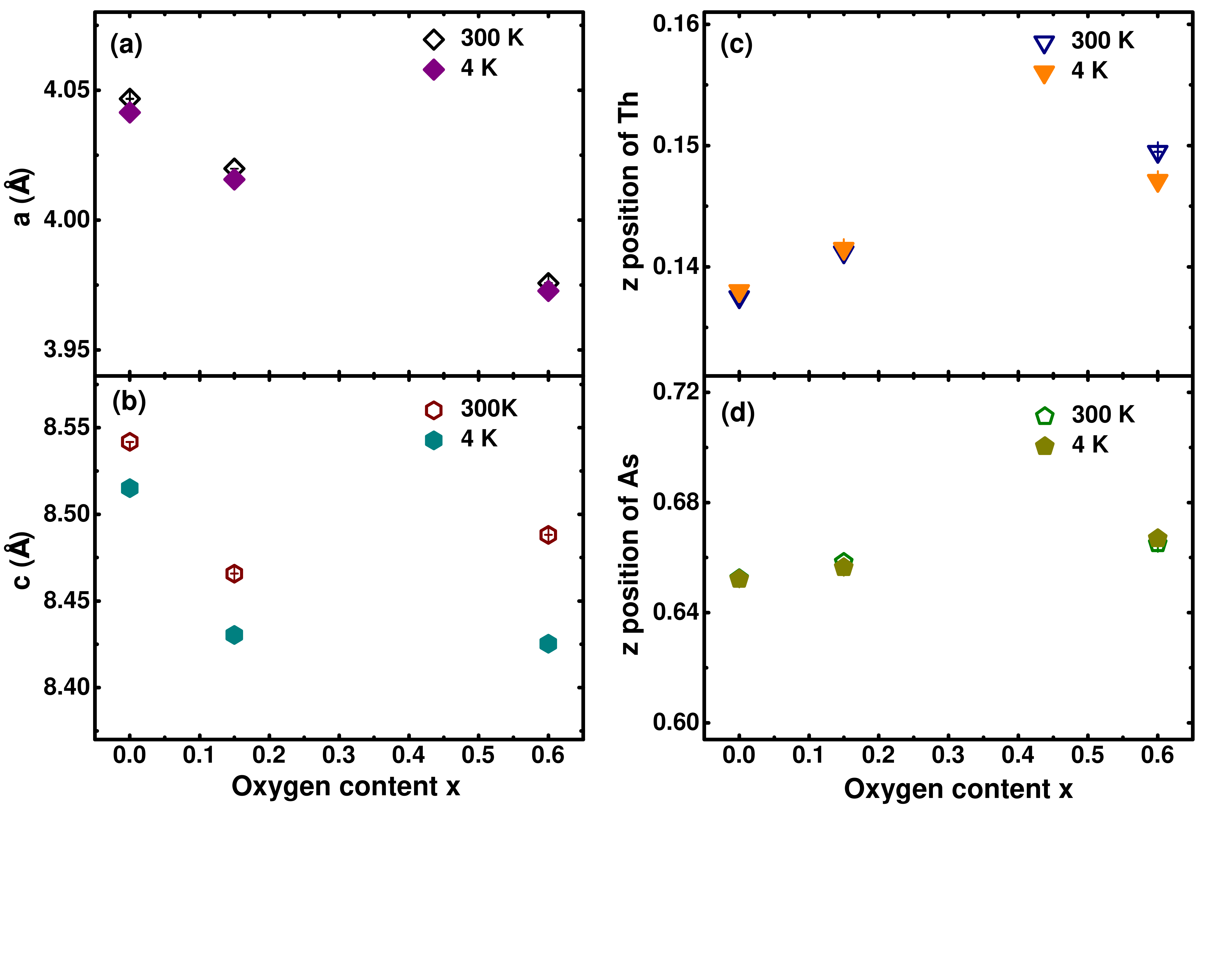}
\caption{Oxygen doping dependence of the (a, b) lattice parameters and (c, d) $z$ position of Th and As in ThFeAsN$_{1-x}$O$_x$ system.
}
\end{figure}

We finally compare the structural parameters for all three compounds measured by neutron scattering: ThFeAsN, ThFeAsN$_{0.85}$O$_{0.15}$ and ThFeAsN$_{0.4}$O$_{0.6}$. Figure 4 summarizes the oxygen doping dependence of the lattice parameters and $z$ position of Th and As. Both $a$ and $c$ decrease upon oxygen doping at low temperature, while $c$ recovers a little for $x=0.6$ at room temperature (Fig. 4 (a) and (b)), suggesting the volume of unit cell is compressed by chemical doping. The $z$ position of As are almost the same for all three compounds, but the $z$ position of Th obviously increases upon oxygen doping, suggesting the substitution of nitrogen by oxygen mainly affects the [Th$_2$N$_2$] layer. By carefully comparing the results in Table.1 and Table.2 with ThFeAsN data \cite{hcmao2017}, a systematic distortion of the FeAs$_4$ tetrahedron is also found, where $\alpha$$_{As-Fe-As}$ and $\beta$$_{As-Fe-As}$ show different behavior when increasing oxygen doping. The microscopic distortion of the lattice structure rather than the phase transition can be explained by uniaxial chemical pressure, which may give result in the two separated superconducting regions in combination from the doping effect \cite{bzli2018}.

In summary, we have carried out neutron diffraction experiments on non-superconducting phases of ThFeAsN$_{1-x}$O$_x$ with $x=0.15, 0.6$. None of them shows any magnetic orders down to 4 K. Therefore, despite the two superconducting phases with different optimal $T_c$ similar to other 1111-type iron-based superconductors, this oxygen doped nitride iron-based superconductor may not have magnetically ordered parent compounds. Further inelastic neutron scattering experiments are highly desired to measure the spin dynamics and check whether it interplays with superconductivity in this nitride family.

\begin{center}
{\bf Acknowledgements}
\end{center}
This work is supported by the Strategic Priority Research Program of  Chinese Academy of Sciences (XDB07020300, XDB25000000), the Ministry of Science and Technology of China (Nos. 2017YFA0303103, 2017YFA0302903, 2016YFA0300502),  the National Natural Science Foundation of China (Nos. 11374011, 11504347, 11304183, 11674406 and 11822411), and the Youth Innovation Promotion Association of CAS (No. 2016004). S. Li and H. Luo acknowledge the project supported by the Key Laboratory of Neutron Physics (NPL), CAEP (No. 2015AB03). B. Hu acknowledges the support of the Science Challenge Project (No. TZ2016004). The authors are grateful for the help on data analysis from Qingzhen Huang at NIST, USA.

\end{document}